# Comment on "Low-temperature transport properties of non-stoichiometric $La_{0.95-x}Sr_xMnO_3$"

E Rozenberg[1] and M Auslender[2]

[1] Department of Physics,
[2] Department of Electrical and Computer Engineering,
Ben-Gurion University of the Negev, POB 653,
8405 Beer-Sheba, Israel

**Abstract**
In a recent paper (Michalopolou A., Syskakis E. and Papastaikoudis C., 2001 *J. Phys.: Cond. Matter*, **13** 11615) A. Michalopolou et al reported the measurements of electrical resistivity and specific heat at zero magnetic field carried out on polycrystalline non-stoichiometric $La_{0.95-x}Sr_xMnO_3$ manganites. In particular, authors had attributed the low temperature behavior of resistivity (shallow minimum and slight upturn at lowest temperatures) to 3D electron-electron interaction enhanced by disorder, using results of numerical fittings of resistivity versus temperature dependencies at the interval 4.2 – 40 K. It is shown in this comment that such analysis may be not valid for polycrystalline manganites where relatively strong grain boundary effects might mask weak contribution of quantum effects to low temperature resistivity. The crucial test of applicability of the theory of quantum corrections to conductivity in this case is the resistive measurements under non-zero magnetic field.

A. Michalopoulou et al have presented in their recent paper [1] data on electrical resistivity and specific heat measured at zero external magnetic field (*H*) on polycrystalline non-stoichiometric $La_{0.95-x}Sr_xMnO_3$ manganites in the doping region $0 \leq x \leq 0.3$. Using numerical fittings of resistivity (ρ) vs. temperature (*T*) dependencies at the interval 4.2 – 40 K, authors claim that the low temperature (LT) behavior of resistivity (shallow minimum and slight upturn at lowest *T*) may be account by 3D electron-electron interaction enhanced by disorder [2], as it was suggested previously in [3-5] for polycrystalline manganites. At the same time, applicability of the theory of quantum corrections to conductivity (QCC) [2] was checked in detail for single crystals and ceramics of different doped manganites in our recent papers [6,7]. Let us mark briefly main results reported in [6,7] and compare it with these ones of commented paper [1].

(i) It should be emphasized that QCC theory [2] is a "bulk" one, describing influence of quantum effects such as electron-electron interaction and weak localization on LT conductivity of bulk (single crystalline) ***metals and compounds with metallic properties***. Thus, one should be extremely careful using QCC model for description of LT resistivity of polycrystalline samples.

(ii) In particular, ***formal numerical fitting of zero-field resistivity*** upturn at lowest *T* with ~ $-T^{1/2}$-dependence alone ***is not sufficient*** for single valued verification of applicability of QCC theory [6]. A crucial test of any theoretical model in this case is the influence of non-zero *H* on the LT minimum of ρ.

(iii) It was shown in [6] and tested additionally in [7] that above ***minimum*** (observed at $T_{min}$ ~ 20-30 K) ***is flattened and vanished under moderate external H*** about 1.5 and 10 Tesla in $La_{0.5}Pb_{0.5}MnO_3$ (LPMO) and $La_{0.8}Sr_{0.2}MnO_3$ (LSMO) ceramics, respectively. ***Such behavior strongly contradicts to prediction of QCC***, according to which LT minimum of ρ persists and is affected very weakly by *H* about 10 T [6].

(iv) Alternative ***model of carriers' tunneling*** between antiferromagnetically coupled grains, ***taking into account grain boundary (GB) effects*** in polycrystalline manganites, ***provides a fairly well qualitative description of above*** (point (iii)) ***phenomenon*** [6]. But notable differences are observed in LT conductivity for relatively "poor" LPMO and "good" LSMO ceramics (mean grain size is by order of magnitude smaller and residual ρ is by two orders of magnitude higher in former sample comparing to the latter one). Firstly, critical *H* that vanishes "GB" minimum in LSMO is higher than that for LPMO (point (iii)) and, secondly, ***additional field independent very weak LT minimum of ρ that was previously masked by relatively strong "GB" minimum*** appears at $H \geq 10$ T in LSMO, while it is absent for LPMO upon *H*, higher than the critical value [7].

(v) ***This LT minimum of ρ in LSMO polycrystalline sample, as well as slight H-independent upturn of resistivity observed in single crystal of LSMO at liquid helium temperatures were attributed to bulk-like LT conductivity governed by QCC*** [7].

We contend that above points (i) – (v) are extremely important for following comments of results by A. Michalopoulou et al [1].

(1) In general, A. Michalopoulou et al did not present in commented paper [1] any systematical data on conductive and magnetic properties i.e. Curie points, temperatures of metal-insulator transition etc. of investigated $La_{0.95-x}Sr_xMnO_3$ system. However, it was pointed out that the parent compound $La_{0.95}MnO_3$, as well as $La_{0.85}Sr_{0.1}MnO_3$ are insulators (the latter sample may be attributed to inhomogeneous insulator with phase-separated ground state – see p.p. 11617 and 11620 in [1]). Temperature-dependent percolation of metal-like domains within an insulating matrix may be a possible nature of LT minimum of ρ observed in $La_{0.85}Sr_{0.1}MnO_3$ - Fig. 2a in [1]. It means that QCC theory in principal could not be used for analysis of LT conductivity in $La_{0.85}Sr_{0.1}MnO_3$ - see point (i) and absolutely different from QCC physical mechanisms may govern LT conductivity of this sample (see, for example, [8,9]).

(2) Nevertheless, LT upturns of ρ observed for $La_{0.85}Sr_{0.1}MnO_3$, as well as for $La_{0.75}Sr_{0.2}MnO_3$ and $La_{0.65}Sr_{0.3}MnO_3$ are pretty well fitted by $\sim -T^{1/2}$-dependence – Fig. 2 in [1]. This fact well illustrates and strongly supports point (ii) on insufficiency of only numerical fitting of ρ vs. $T$ dependence at $H = 0$ for arguing on applicability of QCC model.

(3) It is impossible, of course, to disclaim in principal existence of QCC-like contribution to LT conductivity of metallic-like polycrystalline $La_{0.75}Sr_{0.2}MnO_3$ and $La_{0.65}Sr_{0.3}MnO_3$ samples. But, taking into account our previous results [6,7] and very recent data by S. Roy et al [10], it is possible to assume that experimentally observed LT minimum of ρ at $H = 0$ – Figs. 2 b, c in [1] originates from GB effects (points (iii), (iv)) in above ceramics. Such supposition is confirmed by about the same values of $T_{min} \sim 20$ K and LT upturn of ρ ~ 1% (compare Figs. 2 b, c in [1] and data [6,7,10]) observed at $H = 0$ on different polycrystals of doped manganites. It is interesting to note that in insulator-like $La_{0.85}Sr_{0.1}MnO_3$ ceramic such upturn is notably stronger (about 4%) - Fig. 2 a in [1].

(4) The crucial test of the validity of interpretation proposed by A. Michalopoulou et al is measurement of LT ρ vs. $T$ dependencies under external $H$ about some Tesla – point (ii). It may be predicted certainly that GB-like minimum of resistivity will be flattened and vanished by such field [6,7,10] and almost *H*-independent very weak bulk-like minimum will appear (if it exists) [7]. Taking into account relatively low values of residual resistivity of $La_{0.75}Sr_{0.2}MnO_3$ and $La_{0.65}Sr_{0.3}MnO_3$ samples, the temperature of such bulk-like minimum (described by QCC model) may be estimated as $T_{min} \leq 10$ K [7].

(5) Finally, let us note that Matthiessen's rule used in [1] for determination of possible mechanism of inelastic contribution to conductivity at $T > T_{min}$ is valid only for metallic-like systems. Thus, the value obtained for insulating-like $La_{0.85}Sr_{0.1}MnO_3$ must be excluded from Fig. 3 in [1]. The simplest way of choice between usual electron-electron and unconventional Furukawa scatterings is measuring and fitting of resistivity through extended interval of temperature above $T_{min}$ (in commented paper [1] such fittings were done throughout $\Delta T$ about 10 - 20 K only).

To conclude, simplified analysis of experimental data presented by A. Michalopoulou et al in [1] based on fitting of low temperature resistivity versus temperature dependencies measured only at zero magnetic field is absolutely insufficient for verification of applicability of QCC model (as well as in previous analogous attempts [3-5]). At the same time, analysis of specific heat data (typical bulk property) presented in [1] seems enough plausible.